\documentclass[preprint]{jpsj2}
\def\vec#1{\mbox{\boldmath $#1$}}
\newcommand{\degree}{$^{\circ}~$}
\title{Neutron scattering study of Kondo lattice antiferromagnet YbNiSi$_3$}
\author{Yuha \textsc{Kobayashi}$^{1}$\thanks{E-mail address: kobayashi@issp.u-tokyo.ac.jp }, 
Takahiro \textsc{Onimaru}$^{2}$, 
Marcos A. \textsc{Avila}$^{2}$ ,
Kenzo \textsc{Sasai}$^1$,
Minoru \textsc{Soda}$^1$,
Kazuma  \textsc{Hirota}$^{1}$ 
and 
Toshiro \textsc{Takabatake}$^{2}$
}
\inst{$^{1}$ Neutron-Scattering Laboratory, Institute for Solid State Physics, University of Tokyo, Kashiwa 277-8581 \\
$^{2}$ Department of Quantum Matter, ADSM, Hiroshima University, Higashi-Hiroshima 739-8530\\}

\abst{The Kondo lattice antiferromagnet YbNiSi$_3$ was investigated by neutron scattering. The magnetic structure of YbNiSi$_3$ was determined by neutron diffraction on a single-crystalline sample. Inelastic scattering experiments were also performed on a pulverized sample to study the crystalline electric field (CEF) excitations. Two broad CEF excitations were observed, from which the CEF parameters were determined. The temperature dependence of the magnetic susceptibility $\chi$ and the magnetic specific heat $C_\textrm{mag}$ were calculated using the determined CEF model, and compared with previous results. 
}

\kword{YbNiSi$_3$, heavy fermion, magnetic structure, crystalline electric field, neutron scattering}

\begin{document}
\maketitle

\section{Introduction} 


Recently Yb-based compounds have been investigated intensively owing to their variety of unusual physical properties resulting from a competition between the Kondo effect and long-range inter-atomic exchange interactions, together with crystalline electric field (CEF) effects. The Kondo effect leads to a Fermi-liquid state at low temperatures made of strongly correlated heavy quasiparticles with enhanced effective masses, called heavy fermions. On the other hand, long-range inter-atomic exchange interactions tend to yield magnetically ordered states. In Yb-based heavy fermion systems, CEF effects split the Yb ion multiplet and play an important role in the physical properties.\cite{loewenhaupt_fischer} To understand the behaviors of Yb and 4$f$ electron systems, more systematic and in-depth studies are imperative. 

We have studied YbNiSi$_3$ using neutron scattering techniques. YbNiSi$_3$ is an orthorhombic Kondo lattice heavy-fermion compound, which shows an antiferromagnetic state below $T_\textrm{N}~=~5.1$ K\cite{avila}. This transition temperature $T_\textrm{N}~=~5.1$ K is especially high among Yb compounds, indicating a strong magnetic exchange interaction. YbNiSi$_3$ is thus considered a model case of almost +3 valence Yb-based heavy-fermion antiferromagnetic compounds, which can help a full recognition of a competing relation between the magnetic interaction and the Kondo effect. 

The effects of CEF excitation on the magnetic susceptibility $\chi$ was pointed by Avila $et~al$.\cite{avila} The magnetic moments were considered to align along the easy magnetic axis $b$ from magnetic measurements, and the effective magnetic moment was estimated to be smaller than 2~$\mu_\textrm{B}$ in the ordered state\cite{avila}. To advance the understanding of YbNiSi$_3$, we performed neutron scattering experiments in this present work. Elastic-scattering measurements using a single crystalline sample were performed to resolve the exact magnetic structure. In addition, inelastic-scattering measurements using a pulverized sample were carried out to observe the CEF excitation scheme. The CEF model was then applied to analyze and explain some macroscopic physical properties which have already been reported.\cite{avila,bud'ko} We also discuss the ground state of Yb ion with the CEF model analysis, which the multiplet of the ground state has been a subject of a considerable discussion.\cite{avila,bud'ko}

\section{Experiment}
\subsection{Magnetic Structure}

Neutron diffraction measurements were performed on the triple-axis spectrometer PONTA installed by ISSP in the JRR-3 reactor of the Japan Atomic Energy Agency(JAEA). The (002) reflection of pyrolytic graphite (PG) was used to monochromate and analyze the neutron beam, together with a sapphire and two PG filters to eliminate higher order contaminations. The neutron beam energy was fixed at 14.7 meV. The horizontal collimation of $40'$~-~$40'$~-~$40'$~-~$80'$ was employed. The single-crystalline sample, sealed in an Al can with He gas, was mounted in an ILL-type orange cryostat. The sample size was about $3~\times~3.5~\times~0.5~\textrm{mm}^3$.

An $\omega-2\theta$ scan was performed after moving to the peak profile center-of-mass of an $\omega$ scan on each reciprocal lattice point studied, below and above the N\'eel temperature. The scan profiles were fitted with a gaussian function to estimate the integrated intensities. The integrated intensities were calculated by multiplying the full-width at half-maximum (FWHM) by the peak height of each reflection.
The diffraction experiments were first performed on the scattering plane ($0~k~l$) and then changed to the ($h~k~0$) plane.

\subsection{Crystalline Electric Field}
Powder neutron inelastic-scattering measurements were also performed on PONTA operated in the neutron energy-loss configuration. The powder sample was obtained by pulverizing about 1.1~grams of single crystals. The final neutron energy was fixed at 14.7~meV, using a horizontally focusing analyzer (HFA). The PG analyzer pieces of HFA are designed to focus the final neutron beam, by the pieces forming a spherical surface. Five PG pieces are aligned in the horizontal plane and focus the same energy, but different wave vector neutrons. This makes the detected intensities almost five times larger than an ordinary flat analyzer. The scattering intensities of CEF excitations are almost $Q$-independent, thus a powder inelastic scattering with HFA is an efficient way to obtain sufficient intensities from CEF excitation signals. For instance, HFA was used to successfully observe CEF excitations of Ho$_2$Sn$_2$O$_7$\cite{kadowaki}. HFA is used not only for CEF excitation experiments, but also used for phonon\cite{wakimoto,sasai} or spin excitation measurements under magnetic field\cite{kohgi} studies.

The conditions of horizontal collimators were $40'$~-~$80'$~-~$radial$~-~$open$. The $radial$ is a radial collimator, which divides the neutron beam path by neutron absorbing blades. The sample is located on the extension of the blades to optimize the condition of HFA. The Sapphire and PG filters were employed to remove higher order contaminations. The energy resolution was measured using a vanadium standard sample as $\Delta\hbar\omega~\sim~1.44$~meV at the elastic position, which is only slightly broader than the calculated value. The 1.1~g pulverized sample was sealed in an Al can and mounted in an ILL-type orange cryostat. The constant-$Q$ scans were performed at various scattering vectors and temperatures to search the magnetic excitations.

\section{Results}
\subsection{Magnetic Structure}

\begin{figure}[htb]
\begin{center}
\includegraphics*[scale=0.3]{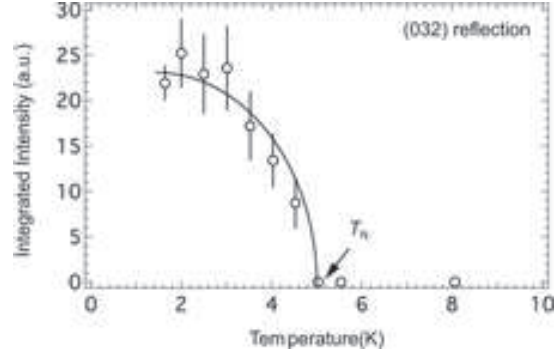}
\caption{The temperature dependence of the (032) magnetic reflection. The solid line is a guide to the eyes.}
\label{f3_1}
\end{center}
\end{figure}

\begin{figure}[htb]
\begin{center}
\includegraphics*[scale=0.3]{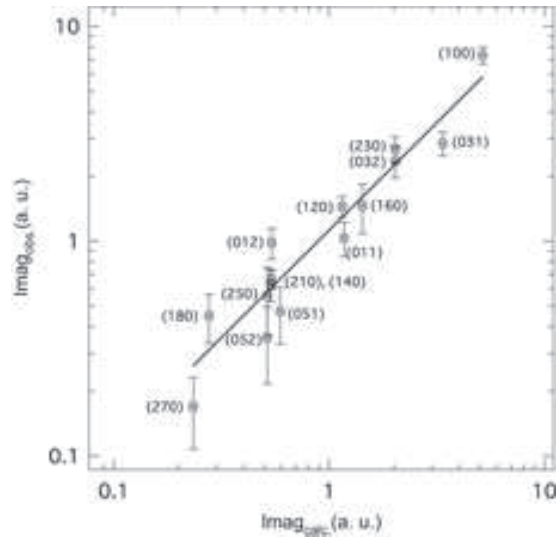}
\caption{The calculated ($Imag_{calc.}$) and observed ($Imag_{obs.}$) intensities of the magnetic reflections. The numbers represent the correspond ($h\:k\:l$) reflections. }
\label{f3_2}
\end{center}
\end{figure}

\begin{figure}[htb]
\begin{center}
\includegraphics*[scale=0.3]{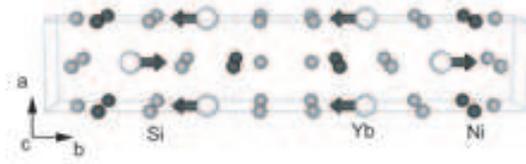}
\caption{The magnetic structure of YbNiSi$_3$. The black arrows denote the directions of the magnetic moments. The Yb moments align ferromagnetically in the $bc$ plane, and the ferromagnetic layers stack antiferromagnetically. White, gray and black circles represent Yb, Si and Ni atoms, respectively.}
\label{f3_3}
\end{center}
\end{figure}

The temperature dependence of the (032) magnetic reflection is shown in Fig.~\ref{f3_1}. The N\'eel temperature is about 5~K, consistent with the previous reports.\cite{avila,bud'ko} The magnetic structure, the propagation vector and the effective magnetic moments of YbNiSi$_3$ were determined by comparing the calculated intensities of several possible models with the measured intensities of the magnetic reflections. The magnetic form factor of Yb$^{3+}$ was used for these calculations.\cite{int_table} The comparison between the calculated and the observed intensities of magnetic reflections is shown in Fig.~\ref{f3_2}. The magnetic reflections were measured at 1.8~K on fifteen reciprocal points which are the forbidden positions of the $Cmmm$ space group. The nuclear reflection data were used for normalization, where the reflections were measured  well above $T_\textrm{N}$. Figure~\ref{f3_3} represents the revealed magnetic structure of YbNiSi$_3$. The black arrows denote the directions of the magnetic moments.

The Yb moments are aligned ferromagnetically in the $bc$ plane and the ferromagnetic layers stacked antiferromagnetically. This magnetic structure is completely different from that of the Ce-based compound CeNiGe$_3$,which has the same crystal-structure.\cite{durivault} The propagation vector was determined to be $\vec{k}~=~(1,~0,~0)$. The effective magnetic moment of YbNiSi$_3$ was determined at $1.1~\pm~0.2~\mu_\textrm{B}$. On the other hand, the magnetization measurements at 2~K shows that the saturated moment is below $2~\mu_\textrm{B}/$f.u.

Very weak gaussian-shaped reflections were also observed at two forbidden positions above $T_\textrm{N}$. The peak height was about 0.6 cps, while the weakest ``not-forbidden'' nuclear reflection was $\sim$ 1 cps. The reflections at the forbidden positions are assumed to be the results of twinning or multiple scattering. If these ``forbidden'' reflections are caused by twinning, the integrated intensities are so weak that the percentage of the twinning should be less than 1\%. The possibility of multiple scattering has not been confirmed by the experiment. The single crystalline YbNiSi$_3$ was so small that it took 15 minutes to measure the (020) reflection, a moderately intense fundamental Bragg peak. Although it would be useful to check for the possibility of multiple scattering experimentally, there was not enough time to perform additional scans. Those very weak forbidden nuclear reflections were thus neglected when resolving the magnetic structure.

\subsection{Crystalline Electric Field}
\begin{figure}[htb]
\begin{center}
\includegraphics*[scale=0.3]{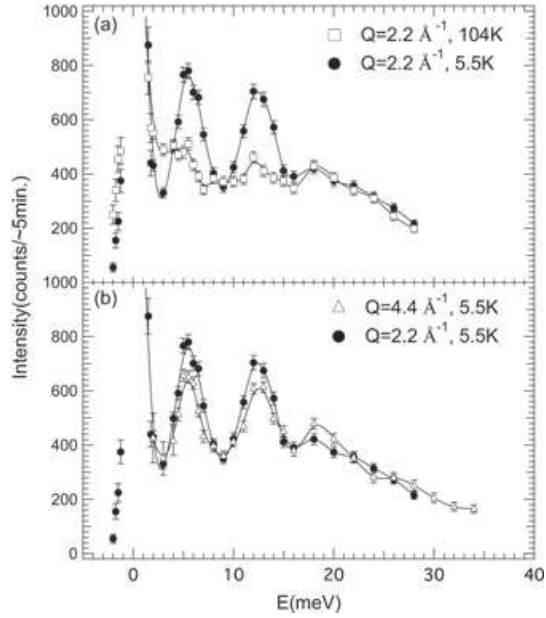}
\caption{The results of the inelastic neutron scattering experiments. (a) The temperature dependence. The scattering vector was fixed at $Q~=~2.2$~\AA$^{-1}$. (b) The $Q$-dependence. The measurements were performed at $T~=~5.5$~K (just above $T_\textrm{N}$). The solid lines are guides to the eyes.}
\label{f3_4}
\end{center}
\end{figure}

Inelastic neutron scattering measurements were performed at $Q~=~2.2$ and 4.4~\AA$^{-1}$, $T~=~5.5~$~K (just above $T_\textrm{N}$) and $T~=~104$~K. Figure~\ref{f3_4}(a) represents the $T$-dependence for $Q~=~2.2~$\AA$^{-1}$ and Fig.~\ref{f3_4}(b) shows the $Q$-dependence  at $T~=~5.5$~K (just above $T_\textrm{N}$) of the inelastic neutron scattering intensities, which were measured using 5G-PONTA. Three inelastic peaks were  observed at ${\Delta}E~=~5.5$, 12 and 18~meV. The peak heights at 5.5 and 12~meV are small at high-$Q$ and high-$T$. Moreover, the $Q$ dependence of the peak heights at 5.5 and 12 meV match that of the square of the Yb$^{3+}$ form factor reasonably\cite{int_table}. Thus we have confirmed that the peaks observed at 5.5 and 12 meV are indeed magnetic excitations. The anomaly at 18 meV is strong at high-$Q$, while there is no clear temperature dependence. Though there might be a contribution of phonon scattering from the Al can\cite{c3} or the sample itself, more detailed experiments are needed to decide the origin of this anomaly. Quasielastic scattering, which is seldom observed in the valence fluctuating systems, were not observed nor resolved. At low temperatures, the quasielastic line width has a positive correlation with the Kondo temperature.\cite{loewenhaupt_fischer} Thus the absence of quasielastic scattering indicates that Kondo temperature of this system is considerably small. 

The measured line width of CEF excitation peaks were considerably broader than the calculated resolution width. When the peaks at 5.5 and 12 meV were fitted with gaussian peaks under the assumption of zero back ground, the line widths (FWHM) were $4.1~\pm~0.2$ and $5.5~\pm~0.4$ meV, respectively. The corresponding calculated resolution widths are 2.2 and 2.8 meV.  
Sometimes Ce- and Yb-based compounds show broad CEF excitations.\cite{boothroyd, donni, so, ivanshin, loong} Usually the hybridization of localized 4$f$ and conduction or valence electrons is employed to explain the broadening, while a CEF-phonon coupling was used to explain the broadening for YbPO$_4$\cite{loong}. In addition, for the case of YbP\cite{donni2}, the broadening was explained by the P-deficiency which changes the local symmetry. No indication of  CEF-phonon coupling nor distortion of Ni or Si were observed in the case of YbNiSi$_3$. Therefore, the hybridization-scenario seems most favorable. However, we need more intensive experiment to conclude which scenario can explain the YbNiSi$_3$ case the best. 

\section{Discussion}
\subsection{Magnetic Structure}
Let us discuss the possibility that the Yb magnetic moments cant toward the $a$ or $c$ axis. The related magnetic reflections would not be observed unless the magnetic moments cant more than 35\degree in our experimental condition, mainly due to the smallness of the single crystal. However, owing to the representational analysis of the space group $Cmmm$ with the propagation vector $\vec{k}~=~(1,~0,~0)$, the magnetic structure is determined absolutely in the framework of the Landau theory. For a second-order transition, the ordering transition can involve only one irreducible representation (IR). The only IR in this case corresponds to the model depicted in Fig.~\ref{f3_3}. In the case of YbNiSi$_3$, the magnetic ordering transition is considered to be second order, and no ferromagnetic properties were observed.\cite{avila,bud'ko} Therefore, we have concluded that the magnetic moments do not cant toward any axes from the $b$ axis. This representational analysis calculation was performed using the SARA$h$-Representational Analysis\cite{c4}.

\subsection{Crystalline Electric Field}
\begin{figure}[htb]
\begin{center}
\includegraphics*[scale=0.25]{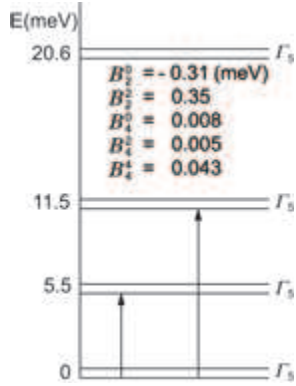}
\caption{The level scheme of YbNiSi$_3$. CEF parameters are also presented. The black arrows indicate the excitations observed by the neutron inelastic scattering experiments.}
\label{f3_5}
\end{center}
\end{figure}

The Stevens' equivalent operator method was used to evaluate the CEF excitation. We have assumed a Kramers doublet as the ground state, though it is a subject of a considerable discussion.\cite{avila,bud'ko}  In the orthorhombic symmetry, the CEF hamiltonian is expressed by 6th degree polynomial. We used the terms up to 4th degree were used for the analysis of the neutron spectrum. That is because it was impossible to reproduce the results of inelastic scattering experiments only by 2nd degree terms, while using the terms up to 6th degree made the analysis confusing. 
The CEF hamiltonian which was used to analyze the inelastic scattering is
\begin{equation}
 \mathcal{H}_{\textrm{CEF}}~=~B^0_2O^0_2~+~B^2_2O^2_2~+~B^0_4O^0_4~+~B^2_4O^2_4~+~B^4_4O^4_4,
\end{equation}

\noindent where $B^i_j$s and $O^i_j$s are the parameters and the Stevens' operators, respectively.\cite{hutchings} The $J~=~\frac{7}{2}$ multiplet of Yb$^{3+}$ ion splits into four $\Gamma_5$ Kramers doublets under this Hamiltonian. On the other hand, we observed only two CEF excitations. Therefore, there might be one more CEF excitations that was not observed nor resolved in the present neutron inelastic experiments. The derived CEF parameters are listed with the corresponding energy diagram in Fig.~\ref{f3_5}. The sign of $B^2_0$ was determined negative from the temperature dependence of magnetic susceptibility $\chi$.\cite{avila,c5,c6} The energy diagram was calculated using the CEF parameters. The black arrows denote the excitations that were observed clearly by the inelastic neutron experiments. The highest excitation state in Fig.~\ref{f3_5} was the result of calculation by the determined CEF parameters.

\begin{figure}[hbt]
\begin{center}
\includegraphics*[scale=0.3]{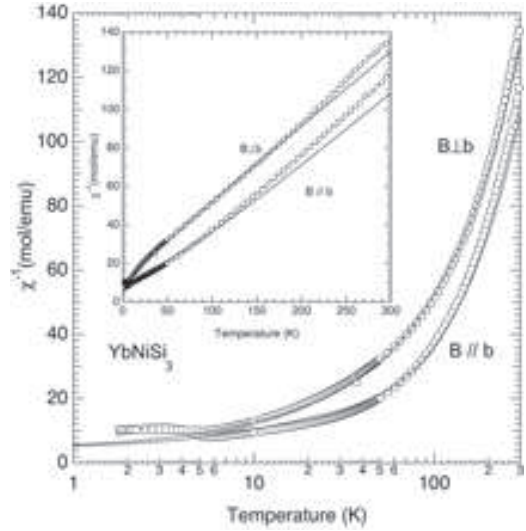}
\caption{The calculated and observed temperature dependence of the magnetic susceptibility against log$-T$. The plots against linear$-T$ is presented in the inset. The solid lines are the calculated results. The open circles are the measured values.\cite{avila}}
\label{f3_6}
\end{center}
\end{figure}

Though it is hard to determine these parameters absolutely, the physical properties can be explained by these determined parameters satisfactorily. The temperature dependence of  $\chi^{-1}$, which was calculated using those determined CEF parameters in the mean-field approximation, is presented in Fig.~\ref{f3_6} along with the measured value of $\chi^{-1}$ against log$-T$.\cite{avila} The calculated and the measured $\chi^{-1}$ against linear$-T$ is plotted in the inset of Fig.~\ref{f3_6}. In the mean-field approximation, $\chi^{-1}_\alpha$ is expressed as $\chi^{-1}_\alpha=\chi^{-1}_{CEF,\alpha}~-~\lambda_\alpha$, where $\lambda_\alpha$ is a molecular-field constant\cite{c7}. The value $\lambda_{ac}~=~\lambda_{b}~=~-~4.5$ mol/emu was used. Although the calculated values were slightly different from the observed data in totality, the magnetic susceptibility of YbNiSi$_3$ above $T_\textrm{N}~=~5~$K is semi-quantitatively explained by this model. For more complete discussion, the 6th degree terms must be included in the CEF hamiltoninan.

\begin{figure}[hbt]
\begin{center}
\includegraphics*[scale=0.3]{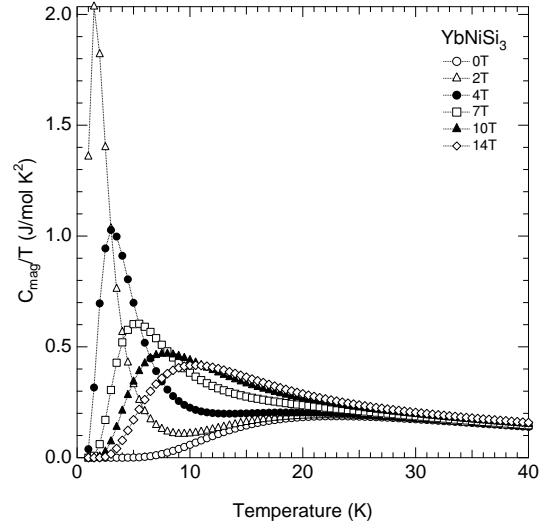}
\caption{The calculated temperature dependences of the specific heat for various applied magnetic fields. Dashed lines are guides to the eyes.}
\label{f3_7}
\end{center}
\end{figure}

The magnetic contribution of specific heat $C_\textrm{mag}$ in an applied magnetic field along the $b$ axis was calculated using thus revealed CEF model. The results are shown in Fig.~\ref{f3_7}. At $zero$ magnetic field, a broad peak exists at 20~K, which is considered a Schottky-type contribution. This broad peak was not observed in the previous report.\cite{avila,bud'ko} The reason why the peak has not been observed is considered to be the results of the release of entropy by the magnetic order transition. Moreover, there is a possibility of the contribution of the lattice specific heat, which might make it difficult to resolve the broad peak. When a magnetic field is applied, a sharp peak appears at $B~=~2$~T. A possible explanation for this behavior is that the lower lying doublet splits by the applied magnetic field. The sharp peak at 2~K becomes broader and moves up to higher temperature with increasing the applied field. Above the N\'eel temperature, these broadenings and movings are qualitatively consistent with the experimental observation.\cite{bud'ko} The assumption of the Kramers doublet ground state is responsible for these results of calculation.


\section*{Summary}
We studied almost +3 heavy fermion antiferromagnet YbNiSi$_3$ by means of neutron scattering techniques. We succeeded in determining the magnetic structure and the effective magnetic moment. Moreover, we observed two broad inelastic magnetic peaks, which are attributed to CEF excitations. The physical properties of YbNiSi$_3$ were reproduced reasonably well by calculation using the CEF level scheme derived from the neutron inelastic scattering.

\section*{Acknowledgments}
We thank Dr. Matsuura for his help in neutron scattering experiments, and Prof.~Sato for useful discussions and advices on the data analysis. This work was performed using 5G-PONTA of the Institute for Solid State Physics, the University of Tokyo. The work at University of Tokyo was supported by Grants in Aid for Scientific Research on Priority Areas ``Novel States of Matter Induced by Frustration'' (Grant No. 19052002). The work at Hiroshima University was supported by Grants in Aid for Scientific Research (A) (Grant No. 18204032) from MEXT, Japan.

\end{document}